\documentclass[aps,twocolumn,tightenlines]{revtex4}

\begin{document}
\title{Anomalous cooling of the parallel velocity in seeded beams}

\author{A. Miffre, M. Jacquey, M. B\"uchner, G. Tr\'enec and J. Vigu\'e}
\address{ Laboratoire Collisions Agr\'egats R\'eactivit\'e -IRSAMC
\\Universit\'e Paul Sabatier and CNRS UMR 5589
 118, Route de Narbonne 31062 Toulouse Cedex, France
\\ e-mail:~{\tt jacques.vigue@irsamc.ups-tlse.fr}}

\date{\today}

\begin{abstract}
We have measured the parallel velocity distribution of a lithium
supersonic beam produced by seeding lithium in argon. The parallel
temperature for lithium is considerably lower than the calculated
parallel temperature of the argon carrier gas. We have extended
the theory of supersonic cooling to calculate the parallel
temperature of the seeded gas, in the limit of high dilution. The
theoretical result thus obtained is in good agreement with our
observations.

\end{abstract}
\maketitle


\section{Introduction}

We have observed an unexpected cooling effect in an argon
supersonic beam seeded with a small amount of lithium: the lithium
parallel temperature is smaller than the argon parallel
temperature. This effect is surprisingly large, the lithium
parallel temperature being $3$ times smaller than the calculated
argon parallel temperature, whereas the common assumption is that
these two parallel temperatures are roughly equal.

Although surprising, this effect can be explained by an extension
of the theory giving the terminal parallel temperature in
supersonic beams of pure gases. This theory, first developed by
Anderson and Fenn in 1965 \cite{anderson65}, has been completed by
Toennies and Winkelmann \cite{toennies77} in 1977 and Beijerinck
and Verster \cite{beijerinck81} in 1981 (for a brief review, see
ref. \cite{toennies77}). The expansion of binary mixtures was
studied by Anderson and coworkers \cite{anderson67,raghuraman77},
as well as by other authors quoted by D.R. Miller in his review
\cite{miller88}. These studies were devoted to the case of a heavy
species seeded in a light species and the fact that the terminal
parallel temperatures of the two species are not equal was
established very early (see figure 8 of reference
\cite{raghuraman77}).  However, we have not found any previous
theoretical or experimental evidence of a difference in parallel
temperatures as large as the one we report here.

The basic phenomenon is due to the exchange of energy between the
alkali atoms and the rare gas atoms. Two effects are important: i)
the alkali-rare gas interaction has a longer range than the rare
gas-rare gas interactions so that the expansion cooling goes on
for a longer time for the alkali atoms; ii) the mass ratio plays
an important role in the energy exchange between the two species
and, as shown below, the parallel temperature of the seeded gas
(mass $m_2$) is considerably lower than the parallel temperature
of the carrier gas (mass $m_1$) when $m_2 \ll m_1$.

In this paper, we first describe our measurement of the parallel
temperature of lithium seeded in argon. Then, we recall the theory
explaining the final temperature in supersonic beams and we
briefly describe its extension to the case of a gas mixture in the
limit of a high dilution. We thus get differential equations
coupling the parallel and perpendicular temperatures of the seeded
gas to the same quantities of the carrier gas. Finally, we compare
our theoretical result to our experiment as well as to another
evidence of this anomalous cooling effect.

\section{Our measurement}

We have measured the parallel temperature of our lithium beam by
laser induced fluorescence. The fluorescence of the $^7Li$ isotope
($92.5\%$ natural abundance) is excited by a single frequency
laser operating at 671 nm, using the $^2S_{1/2}-^2P_{3/2}$
resonance transition. Two laser beams are used, one near normal
incidence measures the transverse velocity distribution, while the
other one, with an incidence close to $45 ^{\circ}$, is sensitive
to a combination of the parallel and perpendicular velocity
distributions. We use a low laser power density in order to
prevent any saturation broadening. Figure 1 presents a typical
spectrum of the fluorescence as a function of the laser frequency.
Each laser beam excites two resonance lines corresponding to the
two hyperfine levels $F=1$ and $F=2$ of the $^2S_{1/2}$ ground
state, with an hyperfine splitting equal to $803.5$ MHz, while the
very small hyperfine structure of the $^2P_{3/2}$ excited state
cannot be resolved \cite{sansonetti95}.

The distance between the lines corresponding to the same
transition excited by the two laser beams gives access to the mean
velocity of the beam, while the linewidths give access to the
widths of the velocity distributions. The observed perpendicular
velocity distribution depends on the width of the observed part of
the lithium beam and our observations correspond to a
perpendicular temperature in the $0.4-0.7$ K range. As expected,
the parallel velocity distribution is broader than the
perpendicular velocity distribution (in our excitation geometry
with a $45 ^{\circ}$ incidence, the Doppler width is only $0.71$
times the width we would get if the laser and atomic beams were
parallel). The measured parallel temperature is $T_{\|} = 6.6\pm
0.6$ K. Unfortunately, as we cannot measure the parallel
temperature of the argon beam, we must deduce its value, $19.0 \pm
2.7$ K, from the source parameters.

\section{Terminal parallel temperature in supersonic
expansions}

We consider here only the case of monoatomic gases. The starting
point is the Boltzmann equation. In the frame moving with the
hydrodynamic velocity $u$, the velocity distribution is assumed to
be Maxwellian with different parallel and perpendicular
temperatures $T_{\|1}$ and $T_{\bot1}$ :

\begin{eqnarray}
\label{m1} f_1({\mathbf v}) &=& \left( \frac{m_1}{2 \pi k_B
T_{\|1}}\right)^{1/2} \times \frac{m_1}{2 \pi k_B T_{\bot1}}
\nonumber \\ &\times& \exp\left[ -\frac{m_1v_{\|}^2}{2k_B
T_{\|1}}- \frac{m_1 v_{\bot}^2}{2k_B T_{\bot1}}\right]
\end{eqnarray}

\noindent $m_1$ is the atomic mass. We use local cartesian
coordinates, the $x$ and $y$ axis being in the perpendicular
directions and the $z$ axis being in the parallel direction. The
theoretical treatment of Toennies and Winkelmann \cite{toennies77}
(noted below TW) and the one of Beijerinck and Verster
\cite{beijerinck81} (noted below BV) are different but equations
(14),(15b) and (16) of reference \cite{toennies77} are equivalent
to equations (29a,b) and (B6,7) of reference \cite{beijerinck81}.
However, small terms of the order of $S_{\|}^{-2}$, where the
parallel speed ratio $S_{\|}$ is defined by $S_{\|} = \sqrt{m
u^2/(2k_B T_{\|})}$ are neglected in the BV calculations, which
are a good approximation only when the terminal value
$S_{\|\infty}$ of the parallel speed ratio is large. As we are
interested in this case here, we follow the BV calculations
because they are simpler. The parallel and perpendicular
temperatures are given by:

\begin{equation}
\label{m2} \frac{dT_{\bot1}}{dz} = -\frac{2 T_{\bot1}}{z} + F
\mbox{  and  }\frac{dT_{\|1}}{dz} = -  2 F
\end{equation}

\noindent where $F$ is due to the collision-induced energy
transfer from parallel to perpendicular degrees of freedom:

\begin{equation}
\label{m3} F = \frac{n_1(z)}{2k_B u} \int g
\frac{d\sigma_{1,1}(g)}{d\Omega} \Delta E f_1({\mathbf v_1})
f_1({\mathbf v}_2) d^3{\mathbf v}_1 d^3 {\mathbf v}_2 d\Omega
\end{equation}

\noindent ${\mathbf v}_1$ and ${\mathbf v}_2$ are the atom
velocities before the collision and $g= |{\mathbf v}_1 - {\mathbf
v}_2|$ is their relative velocity. The indices $1,1$ recall that
the collision considered here involve two atoms of species $1$.
$\Delta E$ is the energy transferred during one collision from the
parallel degree of freedom to the perpendicular ones; after
averaging over the azimuth of the final relative velocity around
the initial relative velocity, $\Delta E$ is given by :

\begin{equation}
\label{m4}\Delta E = \frac{m}{8} \left[g_x^2 + g_y^2 - 2
g_z^2\right] \left[1 - \cos^2\chi\right]
\end{equation}

\noindent where $\chi$ is the deflection angle. $F$ is well
approximated by a linear function of $(T_{\|1}- T_{\bot1}) $ :

\begin{eqnarray}
\label{m5}F & \approx& \Lambda(z)(T_{\|1}- T_{\bot1}) \mbox{ with
} \nonumber \\ \Lambda(z) &=& 16n_1(z)
\Omega_{1,1}^{\left(2,2\right)}(T_m)/(15 u_{\infty})
\end{eqnarray}

\noindent Here, $\Omega_{1,1}^{\left(l,s\right)}(T_m)$ defined by
reference \cite{hirschfelder54} (see equation 7.4-34) is a thermal
average at the mean temperature $T_m = (T_{\|1} + 2 T_{\bot1})/3$
of the collision integral $Q_{1,1}^{(l)}$ given by:

\begin{equation}
\label{m6} Q_{1,1}^{(l)} = \int \frac{d\sigma_{1,1}(g)}{d\Omega}
(1- \cos^l\chi) d\Omega
\end{equation}

\noindent When quantum effects can be neglected, if the
interaction potential is approximated by a 12-6 Lennard-Jones
potential with a $C_6(1,1)$ long-range coefficient and if the
relative kinetic energy is small with respect to the interaction
potential well depth $\epsilon$, the
$\Omega_{1,1}^{\left(2,2\right)}(T)$ integral is given by
\cite{beijerinck81}:

\begin{equation}
\label{m7} \Omega_{1,1}^{\left(2,2\right)}(T) = 2.99 \left(2k_B
T/m_1\right)^{1/2}(C_6(1,1)/k_B T)^{1/3}
\end{equation}
\noindent Knowing $\Lambda(z)$, which behaves like
$z^{-2}T_m^{1/6}$, the differential equations (\ref{m2}) can be
integrated numerically. The parallel temperature tends toward a
constant while the perpendicular temperature decreases
indefinitely, a situation very far from thermodynamic equilibrium.
The perpendicular temperature cools for geometrical reasons
discussed by reference \cite{toennies77}. This calculation relates
the terminal parallel temperature $T_{\|\infty}$ to the density
$n_0$ and temperature $T_0$ in the source and to the nozzle
diameter $d$. The tradition is to give the terminal value
$S_{\|\infty}$ of the parallel speed ratio:

\begin{equation}
\label{m8}
S_{\|\infty} = A \left[ n_0 d
\left(C_6/k_BT_0\right)^{1/3} \right]^{\alpha}
\end{equation}

\noindent  Using SI units, the various values of $A$ and $\alpha$
are $A= 1.412$ and $\alpha= 0.53$ obtained by TW
\cite{toennies77}, $A = 1.313$ and $\alpha = 0.545$ obtained by BV
\cite{beijerinck81}, $A = 1.782$ and $\alpha = 0.495$ obtained
also by BV from a fit of experimental $S_{\|\infty}$ values for
argon expansions. We will use this semi-empirical form of equation
(\ref{m8}) to calculate the argon parallel temperature and we have
estimated a $7$\% error bar on $S_{\|\infty}$ from the dispersion
of the data set used by BV and by comparison with other
measurements \cite{meyer78}.

\section{Generalization to the case of a mixture of two monoatomic gases}

Our calculation is made in the case of a high dilution, i.e. when
the seeded gas density $n_2$ is considerably smaller than the
carrier gas density $n_1$. We neglect the velocity slip i.e. we
assume the same hydrodynamic velocity $u$ for both species.
Because $n_2 \ll n_1$, the expansion of the carrier gas is not
modified by the presence of the seeded gas and the equations
written above remain valid. For the seeded gas, we write similar
equations, considering only the collisions with atoms of the
carrier gas. The calculation of the variation of the parallel and
perpendicular energies of an atom $2$ during a collision with an
atom $1$ is straightforward but too complex to be detailed here.
In order to linearize the equations, we have remarked that the
products $gQ_{1,2}^{(l)}(g)\propto g^{1/3}$ vary slowly with $g$
and we have treated them as independent of $g$. The values of
these products are then chosen to be coherent with the treatment
of the pure gas case described in part III. We thus get the
differential equations relating $T_{\|2}$ and $T_{\bot2}$:

\begin{eqnarray}
\label{g3} \frac{dT_{\bot2}}{dz} &=& -\frac{2 T_{\bot2}}{z} +
\Lambda(z)\rho_s \frac{m_1}{M} \left[T_{\|av} - T_{\bot av}\right]
\nonumber \\ &-&  4\Lambda(z)\rho_s \rho_o
\frac{\mu}{M}\left[T_{\bot2} -T_{\bot1}\right]
\\ \label{g3a} \frac{dT_{\|2}}{dz} &=& - 2 \Lambda(z)\rho_s \frac{m_1}{M}
\left[T_{\| av} - T_{\bot av}\right] \nonumber \\ &-& 4 \Lambda(z)
\rho_s \rho_o \frac{\mu}{M}\left[T_{\|2} -T_{\|1}\right]
\end{eqnarray}

\noindent with $ T_{\|av} = \beta T_{\|1} + \alpha T_{\|2}$,
$T_{\bot av}= \beta T_{\bot1} +\alpha T_{\bot2}$, $\alpha=m_1/(m_1
+ m_2)$ and $\beta = m_2/(m_1 + m_2)$.  $\rho_s$ is the ratio of
$\Omega_{i,j}^{\left(2,2\right)}$ collision integrals:

\begin{equation}
\label{g4} \rho_s =
\frac{\Omega_{1,2}^{\left(2,2\right)}}{\Omega_{1,1}^{\left(2,2\right)}}
= \left[\frac{C_6(1,2)}{C_6(1,1)}\right]^{1/3} \times
\left[\frac{m_1 + m_2}{2 m_2}\right]^{1/2}
\end{equation}

\noindent $\rho_o$ is the ratio of the $l=1$ and $l=2$
angle-averaged cross-sections $\rho_o
=Q_{1,2}^{(1)}/Q_{1,2}^{(2)}$. Using the low energy deflection
function for a 12-6 Lennard-Jones potential (see appendix A of
\cite{beijerinck81}), we have calculated $\rho_o = 1.32$.

Following BV, we have introduced reduced temperatures and a
reduced $z$-coordinate. Then, the numerical integration of the
coupled equations gives the temperatures represented in figure
\ref{temperatures}. At the end of the expansion, the parallel and
perpendicular temperatures of the seeded gas are lower than the
similar quantities for the carrier gas. It is difficult to have a
very simple explanation of this effect, because of the complexity
of equations (\ref{g3},\ref{g3a}). It is nevertheless clear that
the parallel temperature of species $2$ remains coupled to the
perpendicular temperatures when the parallel temperature of
species $1$ is no more coupled.

The parallel temperature ratio $T_{\|2}/T_{\|1}$ can be
substantially lower than $1$. The minimum value for temperature
$T_{\|2}$ is reached when the right-hand side of equation
(\ref{g3}) vanishes. Neglecting the perpendicular temperatures, we
thus get:

\begin{equation}
\label{g5} \frac{T_{\|2,min}}{T_{\|1\infty}} = \frac{m_2
\rho_o}{m_1 + 2 m_2 \rho_o}
\end{equation}

\noindent This limiting value is reached if the number of $1-2$
collisions is sufficiently larger than the number of $1-1$
collisions, i.e. if the ratio $\rho_s$ is large enough. We have
plotted in figure \ref{temperature ratio} the terminal value of
the ratio $T_{\|2}/T_{\|1}$ for various values of the ratio
$\rho_s$ and for different values of the mass ratio $m_2/m_1$.

\section{Comparison with experimental results}

We may now compare our theoretical and experimental results for
lithium seeded in argon. The argon (from Air Liquide stated purity
$99.999$\%) is further purified by a cartridge (also from Air
Liquide). The lithium pressure is fixed by the temperature of the
back part of the oven at $0.55$ mbar. In our experiment with $p_0
= 333$ millibar, $T_0 = 943$ K and a nozzle diameter $d= 200$
$\mu$m, the parallel temperature of lithium is $T_{\|2\infty} =
6.6\pm 0.6$ K. Using equation (\ref{m8}), we deduce the argon
parallel temperature from our source conditions, $T_{\|1\infty} =
19.0 \pm 2.7 $ K. We thus get a parallel temperature ratio
$T_{\|2\infty}/T_{\|1\infty} = 0.35\pm 0.08 $. This value is in
good agreement with our theoretical result
$T_{\|2\infty}/T_{\|1\infty} = 0.31$ obtained with $\rho_s = 2.55$
deduced from the $C_6$ values of argon-argon and lithium-argon
interaction \cite{tang76}.

A similar beam with sodium seeded in argon was built by D.
Pritchard and coworkers \cite{schmiedmayer97,ekstrom95} with a
source temperature near $1000$ K, an argon pressure up to $p_0 =3$
bar and a nozzle diameter $d = 70$ $\mu$m. The measured velocity
$u= 1040 \pm 2$ m/s corresponds to $T_0= 1039$ K and the rms
velocity width, deduced from the atomic diffraction pattern, was
found equal to $3.7 \pm 0.4$\% , corresponding to $T_{\|2\infty} =
4.1 \pm 0.9$ K. From the calculated argon parallel temperature
$T_{\|1} = 7.7\pm 1.1$ K corresponding to the largest pressure, we
get the parallel temperature ratio $T_{\|2\infty}/T_{\|1\infty} =
0.53 \pm 0.19$, in good agreement with our theoretical result is
$T_{\|2\infty}/T_{\|1\infty} = 0.61$, using the $C_6$ values from
reference \cite{tang76} ($\rho_s=1.67$).

\section{Conclusion}

In this letter, we have described a measurement of the parallel
velocity distribution of a supersonic beam of lithium seeded in
argon, with a high dilution. The measured lithium parallel
temperature $6.6\pm 0.6$ K is considerably lower than the
calculated parallel temperature of the argon beam near $19.0\pm
2.7$ K.

We have extended the theory of the velocity distribution in
supersonic beams to a mixture of monoatomic gases, in the high
dilution limit. We have established differential equations
describing the parallel and perpendicular temperatures of the
seeded gas; these equations explain the observed effect as a
consequence of several circumstances: role of the mass ratio in
energy exchange during collisions of the two species, longer range
of the lithium-argon interaction as compared to argon-argon
interaction. We have deduced the lowest possible value of the
temperature ratio $T_{\|2\infty}/T_{\|1\infty}$. This ratio may be
quite low when a light species is seeded in a heavy species with
$m_2 \ll m_1$.

We have compared our theory with experimental results
corresponding to lithium seeded in argon (our experiment) and
sodium seeded in argon (an experiment done by D. Pritchard and
coworkers). In both cases the experimental and theoretical values
of the ratio of the parallel temperatures are in good agreement,
but the error bars on the experimental values are large. A
simultaneous measurement of the parallel temperatures of the
carrier and seeded gases would be useful to reduce these error
bars, thus providing a better test of the approximations done in
our theoretical analysis.

\section{Acknowledgements}

We thank J.P. Toennies  for his help and interest in this question
and P. A. Skovorodko for helpful discussions. We also thank Ph.
Dugourd and M. Broyer for advice concerning the design of the oven
of our lithium supersonic beam. We thank R\'egion Midi
Pyr\'en\'ees for financial support.


\begin{figure}
\caption{\label{fluo} Laser induced fluorescence signal as a
function of the laser frequency: the dots represent the
experimental data and the full curves are the Gaussian fits to
each line. The Fabry-Perot used for calibration has a free
spectral range equal to $251.4 \pm 0.5$ MHz. The fluorescence
peaks are labeled by the ground state $F$ value ($F = 1$ or $2$)
and by a letter corresponding to the laser beam: A for the beam
near $45^{\circ}$ incidence, B for the beam near normal
incidence.}
\end{figure}

\begin{figure}
\caption{\label{temperatures} The reduced temperatures $\tau =
\Xi^{-9/11} T/T_0$ are plotted as a function of the reduced
z-coordinate $\zeta = 2.48 \Xi^{12/11} z/d$ with $\Xi =0.813 n_0 d
\left(C_6/k_BT_0\right)^{1/3}$ (to give an example, for our argon
beam seeded with lithium, $\Xi = 32.4$). This plot, which is
identical to figure 5 of Beijerinck and Verster $[3]$ for
$\tau_{\|1}$ and $\tau_{\bot1}$, shows the anomalous cooling
effect for species $2$. We have used $\rho_o= 1.32$ and $\rho_s =
2.55$ corresponding to lithium seeded in argon.}
\end{figure}

\begin{figure}
\caption{\label{temperature ratio} The calculated parallel
temperature ratio $T_{\|2\infty}/T_{\|1\infty}$ is plotted as a
function of the mass ratio $m_2/m_1$ (spanning the range from
hydrogen to rubidium seeded in argon) for several values of the
ratio $\rho_s$, with $\rho_o$ fixed, $\rho_o = 1.32$. From bottom
to top, $\rho_s=\infty$, $\rho_s=2.55$ (lithium case), $\rho_s
=1.67$ (sodium case). The experimental points corresponding to
lithium and sodium seeded in argon are represented by stars.}
\end{figure}

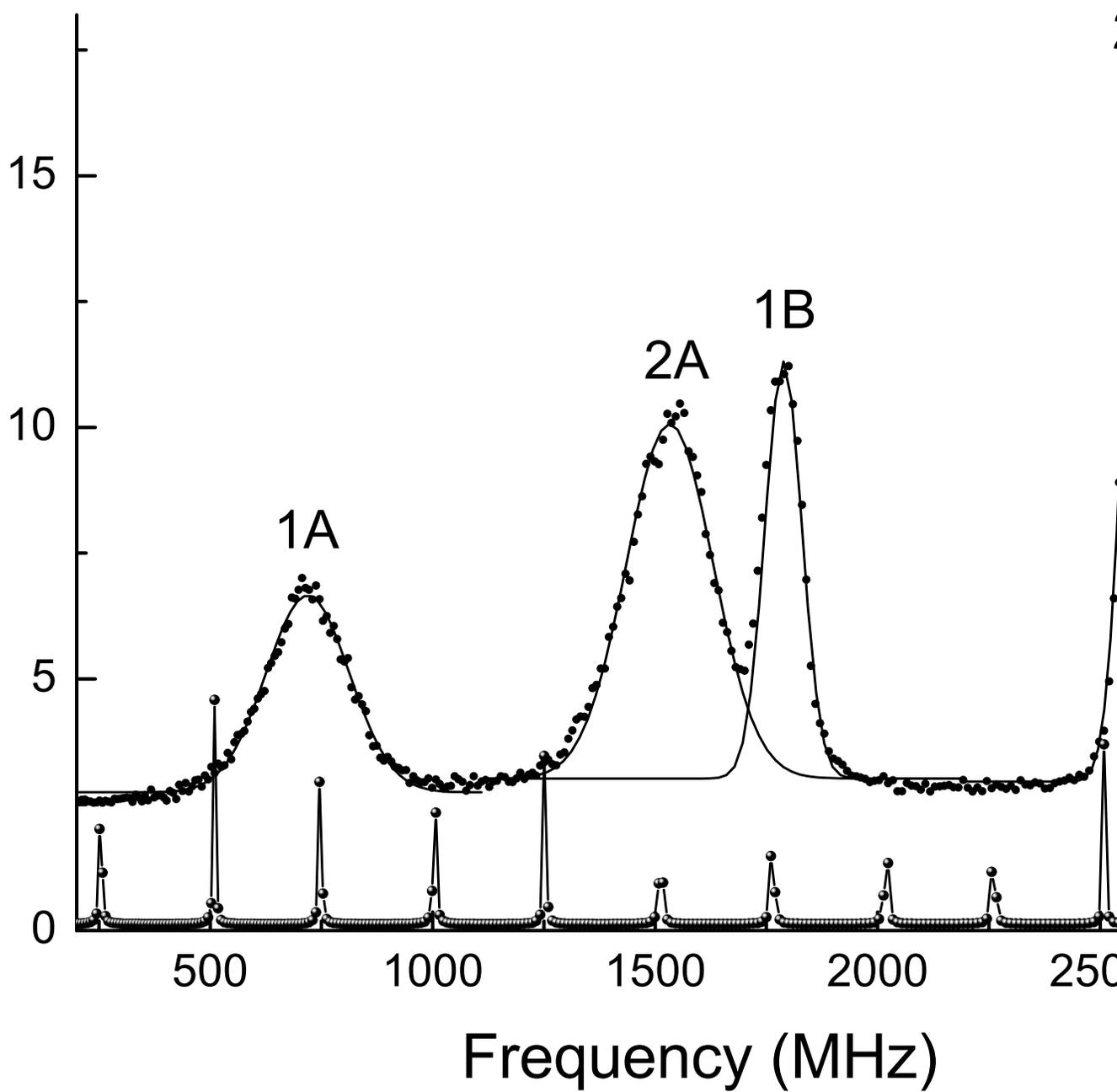
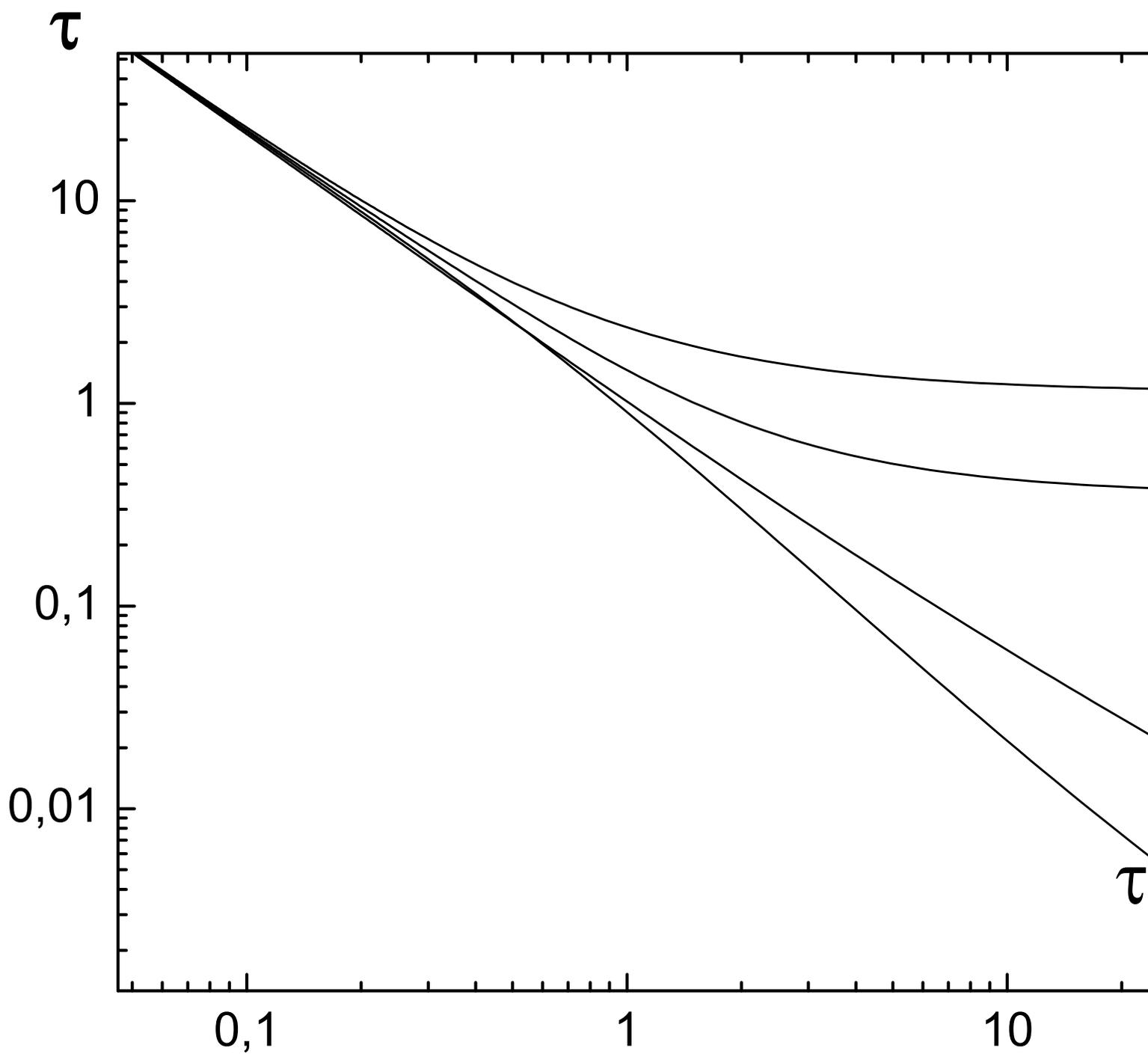
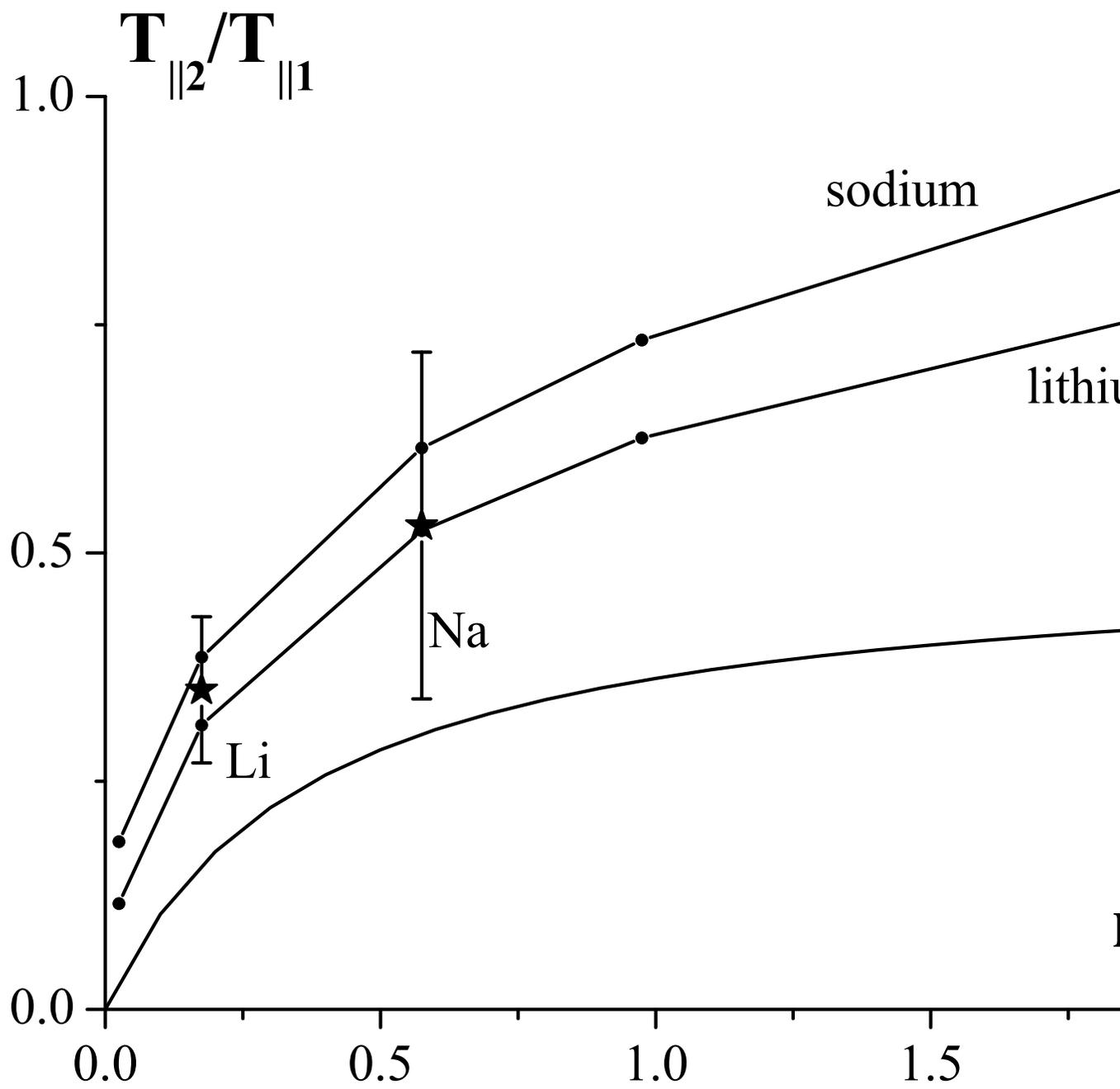
\end{document}